\documentclass[sagev,times,english,greek]{sagej}

\usepackage{moreverb,url}
\usepackage{babel}
\usepackage{setspace}
\newcommand{\gtap}{\mathrel{\hbox{\rlap{\lower.55ex \hbox {$\sim$}}
                   \kern-.3em \raise.4ex \hbox{$>$}}}}
\newcommand{\ltap}{\mathrel{\hbox{\rlap{\lower.55ex \hbox {$\sim$}}
                   \kern-.3em \raise.4ex \hbox{$<$}}}}
\newcommand{\izzit}{\mathrel{\hbox{\rlap{\lower.05ex \hbox {$=$}}
                   \kern-.23em \raise1.10ex \hbox{\footnotesize ?}}}}
\usepackage{color}

\newcommand{\omdac}{\~w\hspace{-0.4em}|\hspace{0.3em}}
\newcommand{\omdacdot}{\~w\hspace{-0.4em}|\hspace{0.3em}.}
\newcommand{\iotdac}{\~{>{\hspace*{-0.2em}}i}}
\newcommand{\etadac}{\~{>{\hspace*{-0.3em}}h}}
\def\farcm{\hbox{$.\mkern-4mu^\prime$}}

\begin{document}
\selectlanguage{english}

\runninghead{Verbunt and van der Sluys}

\title{Why Halley did not discover proper motion and why Cassini did}

\author{Frank Verbunt\affilnum{1,2} and Marc van der Sluys\affilnum{1}}

\affiliation{\affilnum{1}Department of Astronomy / IMAPP, Radboud University
  Nijmegen, the Netherlands\\
  \affilnum{2}SRON Netherlands Institute for Space Research, Utrecht, the Netherlands}

\corrauth{Frank Verbunt,
Department of Astronomy / IMAPP,
Radboud University,
PO Box 9500,
6500 GL Nijmegen, The Netherlands}

\email{f.verbunt@astro.ru.nl}

\begin{abstract}In 1717 Halley compared contemporaneous measurements
of the latitudes of four stars with earlier measurements by ancient
Greek astronomers and by Brahe, and from the differences concluded
that these four stars showed proper motion. An analysis with modern
methods shows that the data used by Halley do not contain
significant evidence for proper motion. What Halley found are the measurement errors
of Ptolemaios and Brahe. Halley further argued that the  occultation of Aldebaran 
by the Moon on 11 March 509 in Athens confirmed the change in
latitude of Aldebaran. In fact, however, the relevant  observation was almost
certainly made in Alexandria where Aldebaran was not occulted. 
By carefully considering measurement errors Jacques Cassini showed that
Halley's results from comparison with earlier astronomers were spurious,
a conclusion partially confirmed by various later authors.
Cassini's careful study of the measurements of the latitude of Arcturus
provides the first significant evidence for proper motion.
 \end{abstract}


 \maketitle

\section{Introduction}

The possibility of motion of the stars relative to one another was
raised by Hipparchos, around 130 BC, as we know from the discussion by
Ptolemaios in the first chapter of Book 7 of the
Almagest\cite{toomer98}. Hipparchos argued that the conclusion that
the stars are fixed on the celestial sphere requires proof that they
do not move with respect to one another. He investigated this by
considering configurations of stars, such as alignments, in which it
is easier to detect relative position changes, and by comparing his
data with those of Timocharis, who lived one and a half century
before.  Hipparchos found no evidence for relative motion.  Ptolemaios
affirms this result over the longer time span of about 280 years
separating him from Hipparchos and even more from Timocharis. 
  The configurations and alignments connect stars in the Zodiac to
  those above or below it, since the main goal was to discover whether
  stars outside the Zodiac increase their longitudes at the same rate
  as the stars in the Zodiac. However, Ptolemaios notes explicity that the
  study of these alignments can also uncover displacements of stars
  within the separate parts of the alignments.

There is possibly an echo to this in the statement by
Macrobius\cite{macrobius} who lived around 400 AD, that the stars have
their {\em `own motion'} (suo motu) on top of their motion with the heavenly
sphere.  Macrobius refers to the incredible number of centuries
required for the completion of {\em `one circuit'} (una ambitio) implying
repeated passages in a closed loop.  The main problem in
interpreting Macrobius is that he clearly doesn't know what he is
talking about.  Because of his reference to a closed loop, it is
probable that his statement refers not to proper motion of
individual stars, but to precession.

Brahe\cite{brahe02}  concluded that the ancient measurements are
rather inaccurate. For example, he compares the latitudes of Aldoboram (Aldebaran)
that result from the measurements by Timocharis, Hipparchos
and Ptolemaios, and `{\em cannot but wonder}' about the large
differences between them. He  warned, correctly, that the latitude from 
Ptolemaios is erroneous: `{\em widest of the mark}'.

In 1717 Halley\cite{halley17} published {\em Considerations on the
  change of the latitudes of some of the principal fixt stars}, in
which he ignores the warning by Brahe and
compares contemporaneous measurements with those of Hipparchos and
Ptolemaios for the latitude of four stars, viz.\ Palilicium or the
Bull's Eye (i.e.\ Aldebaran), Sirius, Arcturus and the bright shoulder
of Orion (Betelgeuse).  He also compares contemporaneous measurements
of Sirius with those of Brahe.  On the basis of this he suggested that
all four stars had shown proper motion.

In the same paper, Halley refers to an occultation of Aldebaran
observed on 11 March 509, in or near Athens according to  the 17th century
French astronomer and polymath Isma\"el Boulliau
(Bullialdus\cite{bullialdus45}): `{\em when in the beginning of the
  night the Moon was seen to follow that star very near, and seemed to
  have eclipsed it}'.  Boulliau computes that such an occultation
could not have happened, from which Halley concluded that an actual
occurrence of an occultation was possible only if `{\em the latitude
  of Pallicium were much less than we at this time find it}'.  As we
will see below Halley refers to the absolute value of the latitude,
and implies a proper motion in the southern direction.

By the time of Halley the concept that stars are attached to a
  sphere, or to spheres if they do not all participate in the same
  precessional motion, had been replaced with the idea of stars moving
  in three-dimensional space. Because in the Copernican system the
  daily, yearly and long-term precessional motions of the stars are
  apparent, merely reflecting rotation, revolution and precession of
  the axis of the Earth, there was no need to assume that the stars
  all share the same motions because they are attached to a sphere.  
  With his discovery that planets move in ellipses,
  Kepler was forced to conclude that the planets move freely in space,
  thereby also undermining the concept of a sphere of the stars. It is
  important to note, however, that 18th century astronomers could no more
  determine distances or radial motions of the stars than Hipparchos
  or Ptolemaios, so that technically the challenge of observationally
  determining proper motion is the same in the 18th century as in
  ancient Greece.

How conclusive is the evidence produced by Halley for proper motions?
This was questioned already by Jacques Cassini\cite{cassini38}, the
son of Gian Domenico Cassini, who argued that the measurements by
Hipparchos / Ptolemaios were too inaccurate to be of use for the
determination of proper motion, and who corrected the latitudes given
by Brahe.  In modern times, van de Kamp\cite{vandekamp86} notes that
the proper motions of Aldebaran and Betelgeuse are very small, and
that Halley's results for these stars must be considered
spurious. Even so, the determination of proper motions of Sirius and
Arcturus by Halley is often still considered valid\cite{brandt10}.  With regard
to the occultation, Neugebauer\cite{neugebauer75}, referring to
computations by Stephenson, agrees with Boulliau that no occultation
took place! Notwithstanding these problems, Halley is still generally
credited with the discovery of proper motion.  In this paper we take a
close look at the argumentation by Halley, and at the study by
Cassini.

\begin{table}
\small\sf\centering
\caption{Ecliptic latitudes of the four  stars discussed by Halley in various
  catalogues: Ptolemaios / Hipparchos in a modern edition
  (Toomer\cite{toomer98}) and in an edition by Hudson\cite{hudson12};
  Brahe in the edition by Kepler\cite{kepler27}; and Flamsteed\cite{flamsteed25}.
  For each catalogue the one-sigma error $\sigma_\beta$ in the
  latitude is listed (from Verbunt \&\ van Gent\cite{verbuntgent12},
  \cite{verbuntgent10}, and Lequeux\cite{lequeux14}).
  We list the difference $d_\beta$ between the catalogue latitude 
  $\beta_\mathrm{cat}$ with the correct latitude $\beta_\mathrm{HIP}$ 
  computed for the catalogue equinox  from data obtained with the 
  HIPPARCOS satellite\cite{leeuwen07}: 
  $d_\beta= \beta_\mathrm{cat}-\beta_\mathrm{HIP}$.
  We also list the latitudes measured by Richer and Cassini, and
  from the re-analysis of Brahe's measurements by Cassini\cite{cassini38}. 
  \label{t:latitudes}}
\begin{tabular}{|lrr|rrrr|rrrr|}
\toprule
&& $\sigma_\beta$ &\multicolumn{3}{c}{$\beta$\,(Arcturus)}  &
$d_\beta$&\multicolumn{3}{c}{$\beta$\,(Sirius)}  & $d_\beta$ \\
catalogue & date &($'$)& $^\circ$ & $'$ & $"$ & ($'$)& $^\circ$ & $'$ & $"$ & ($'$) \\
\midrule
Ptolemaios &$-128$  &23& +31 & 30 & & $-41.4$&$-$39 & 10 & & $-2.4$ \\
\, ed.\ Hudson & && +31 & 10 & &$-61.4$ &  $-$39 & 10 & &$-2.4$\\
Brahe & 1601&2&  +32 & 02 & 30 & 1.8& $-$39 & 30 & & 0.8 \\
Flamsteed &1690&0.5&+30 & 57 & 00 &$0.0$ &$-$39 & 32 & 08 &$-0.1$\\
\midrule
Brahe/Cassini & 1584 & &+31 & 00 & 29 & $-$1.0 & $-39$& 32 & 10 &  $-1.5$ \\
Richer & 1672 & & +30 & 57 & 25 & $-0.4$ & $-39$ & 31 & 55 & $-0.1$\\
Cassini & 1738 & & +30 & 55 & 26 & 0.4 & $-39$ & 33 & 00 & (a) \\
\toprule
& & $\sigma_\beta$ &\multicolumn{3}{c}{$\beta$\,(Aldebaran)} & $d_\beta$ & \multicolumn{3}{c}{$\beta$\,(Betelgeuze)}  &$d_\beta$  \\
catalogue & date & ($'$)& $^\circ$ & $'$ & $"$ & ($'$)& $^\circ$ & $'$ & $"$ & ($'$) \\
\midrule
Ptolemaios &$-$128 & 23 & $-$5 & 10 & & 27.0 & $-$17 & 00 && $-41.1$\\
\, ed.\ Hudson & && $-$5 & 30 & &7.0 &$-$17 & 10& & $-51.1$\\
Brahe & 1601 & 2&$-$5 & 31 & &$-1.3$& $-$16 & 06 &&$-1.2$\\
Flamsteed & 1690 &0.5&$-$5 & 29 & 49  & $-0.5$ &$-$16 & 04 & 26 & $-0.3$ \\
\midrule
Brahe/Cassini  &1589 && $-5$ & 30 & 23 & $-0.6$ & & & & \\
Cassini & 1738 & & $-5$ & 29 & 34 & $-0.4$ & & & & \\
\bottomrule
\end{tabular}
\vspace*{0.3cm}

(a) the latitude of Sirius is given by Cassini as `larger by about 
a minute' than found in Flamsteed, Richer and Cassini's reanalysis of Brahe.
\end{table}
\vspace*{0.2cm}

\section{Latitude differences: Halley}

\begin{table}
\small\sf\centering
\caption{Columns 2-5: Difference in latitude found by subtracting from the values 
given by Ptolemaios (left) or Brahe (right) the values found by converting the positions 
from Halley's time to the equinox of the old  catalogue.
  (a) as given by Halley, (b) converting positions from Flamsteed using
  obliquity values used by Halley, (c) idem using correct values for obliquity
  (d) the actual difference $\Delta\beta_\mathrm{HIP}$ between latitudes
  computed for both epochs from HIPPARCOS satellite data.  \label{t:pm}}
\begin{tabular}{|l|rrrr|rrrr|}
\toprule
 & \multicolumn{4}{|c|}{$\Delta\beta(')$ Ptolemaios to Halley} &  \multicolumn{4}{|c|}{$\Delta\beta(')$ Brahe to Halley}\\
star & (a) &  (b) & (c) & (d)  & (a) &  (b) & (c) & (d) \\
\midrule
Aldebaran & 35 &38.6 & 33.4 & 5.9&  & 1.1 & $-0.5$ & 0.3\\
Sirius        & 42  & 44.1 & 36.2& 38.6& 4.5 & 4.6 & 2.8  & 1.9\\
Arcturus   & 33& 28.1 & 27.2  & 68.6 & & 4.6 & 5.2 & 3.3\\ 
Betelgeuse & $\sim-60$ & $-34.1$ & $-41.1$ & $-0.3$ & & 0.9 & $-0.9$&0.0 \\
\bottomrule
\end{tabular}
\end{table}

Precession at constant obliquity leads to an increase of the ecliptic
longitude, but does not affect the latitude. The slow change of
obliquity leads to slow changes in latitude.  In
Table\,\ref{t:latitudes} we collect ecliptic latitudes of Aldebaran,
Sirius, Arcturus and Betelgeuse from the star catalogues of
Ptolemaios, Brahe, and Flamsteed.  The latitudes given by
Ptolemaios (taken from Toomer's edition\cite{toomer98}) differ by much more
from those given by 17th century astronomers than can be explained
with the change in obliquity. In our analysis we assume that Ptolemaios
corrected longitudes determined by Hipparchos, undo his
corrections and use the epoch of Hipparchos, $-128$\cite{verbuntgent12}. 
That Ptolemaios corrected measurements by Hipparchos for precession
rather than make observations himself, was concluded by Brahe; Halley
will have been aware of this. Our results are not affected by this choice.

In his 1712 edition of the star catalogue of Ptolemaios, Hudson\cite{hudson12} 
acknowledges emendations made by Halley. In the same year Halley edited a star
catalogue\cite{anonymous12} based on data from Flamsteed, which he
had surreptitiously obtained in collusion with Newton\cite{westfall80}. This work may
have alerted Halley to large latitude differences.  (The positions of
the four stars in Table\,\ref{t:latitudes} in the 1712 pirate edition are
identical to those in the 1725 edition by Flamsteed\cite{flamsteed25}
himself.)

The proper motions in the latitude direction cannot be derived simply
by dividing the difference between the latitudes for different
catalogues by the time interval, but require correction for the change
in obliquity between the catalogue equinoxes. Halley remarks that the
value for the obliquity used by Brahe is $2\frac{1}{2}$ [sc.\ arcmin]
larger than in Halley's time.  Since Brahe used
$\epsilon=23^031'30''$, this implies that Halley for his own epoch
used $\epsilon=23^\circ29'$, and (since he mentions that the value at
the time of Ptolemaios was 22$'$ larger) $\epsilon=23^\circ51'$ for
150, the epoch he assumed for Ptolemaios.  With these values Halley
makes the conversions of the latitude from the equinox of his time to
that of Ptolemaios, and derives the remaining differences in
latitude. We list his results in Table\,\ref{t:pm}. Halley gives no
information on his sources for the star catalogue by Ptolemaios and
for the contemporaneous catalogue, but it stands to reason that for
the latter he used the data of Flamsteed. From
Table\,\ref{t:latitudes} we see that use of the catalogues of Brahe
would have led to very similar values.  Halley also does not explain
how he computed the precession.

The one-sigma errors $\sigma_\beta$ in latitude and the deviations
$d_\beta $ in latitude by Ptolemaios (Table\,\ref{t:latitudes}) are
comparable in size to the latitude differences $\Delta\beta$ derived
by Halley: in modern parlance only the (very wrong) $\Delta\beta$ of
Betelgeuse is significant at the two-sigma level. The error
distribution of the latitudes in the star catalogue of Ptolemaios has
more large deviations than a Gaussian\cite{verbuntgent12}, hence the
significance of the $\Delta\beta$ values is even less than estimated
from the Gaussian with $\sigma_\beta=23'$. Indeed, three of the four
stars in Table\,\ref{t:latitudes} have $|d_\beta|>23'$. This indicates
that the $\Delta\beta$ values by Halley are spurious, and that the
proximity of the value for Sirius to the correct value is accidental.
An interesting twist is given by the positions assigned to Aldebaran
and Arcturus by Hudson, who acknowledges Halley, in his edition\cite{hudson12} of the star
catalogue of Ptolemaios. These values, which we also list in
Table\,\ref{t:latitudes},  imply much smaller proper
motions. Apparently, Halley changed his mind between 1712 and
1717\ldots

The exercise of comparing contemporaneous latitudes with
those in an old catalogue was repeated by Halley for the catalogue of
Brahe, for Sirius only. 
Halley\cite{halley17} gives two values for $\Delta\beta$ for
Sirius. Taking into account that the catalogue of Brahe has more
large errors than described with a Gaussian\cite{verbuntgent10},
even the larger value of 4.5$'$ is not significant at the two-sigma
level. In addition, Halley allows the possibility that
$\Delta\beta=2'$ (albeit for the somewhat unlikely 
assumption that Brahe ignored atmospheric refraction).

The uncertainty in the obliquity also adds to the errors in the
comparison with Ptolemaios.  At the end of his article
Halley\cite{halley17} expresses some doubt that the value for the
obliquity at the time of Hipparchos to Ptolemaios was indeed 22$'$
larger than in his own time. This doubt is justified, as the correct
difference in obliquity is 15$'$ and 13$'$ for the epochs of Hipparchos
and Ptolemaios, respectively (Seidelman\cite{seidelman92}).  The
deviations in latitude due to the uncertainty in $\epsilon$ are much
smaller than the typical uncertainty $\sigma_\beta$ in the star
catalogue of Ptolemaios. We return to this below.

\begin{figure}
\center{\includegraphics[width=0.56\columnwidth]{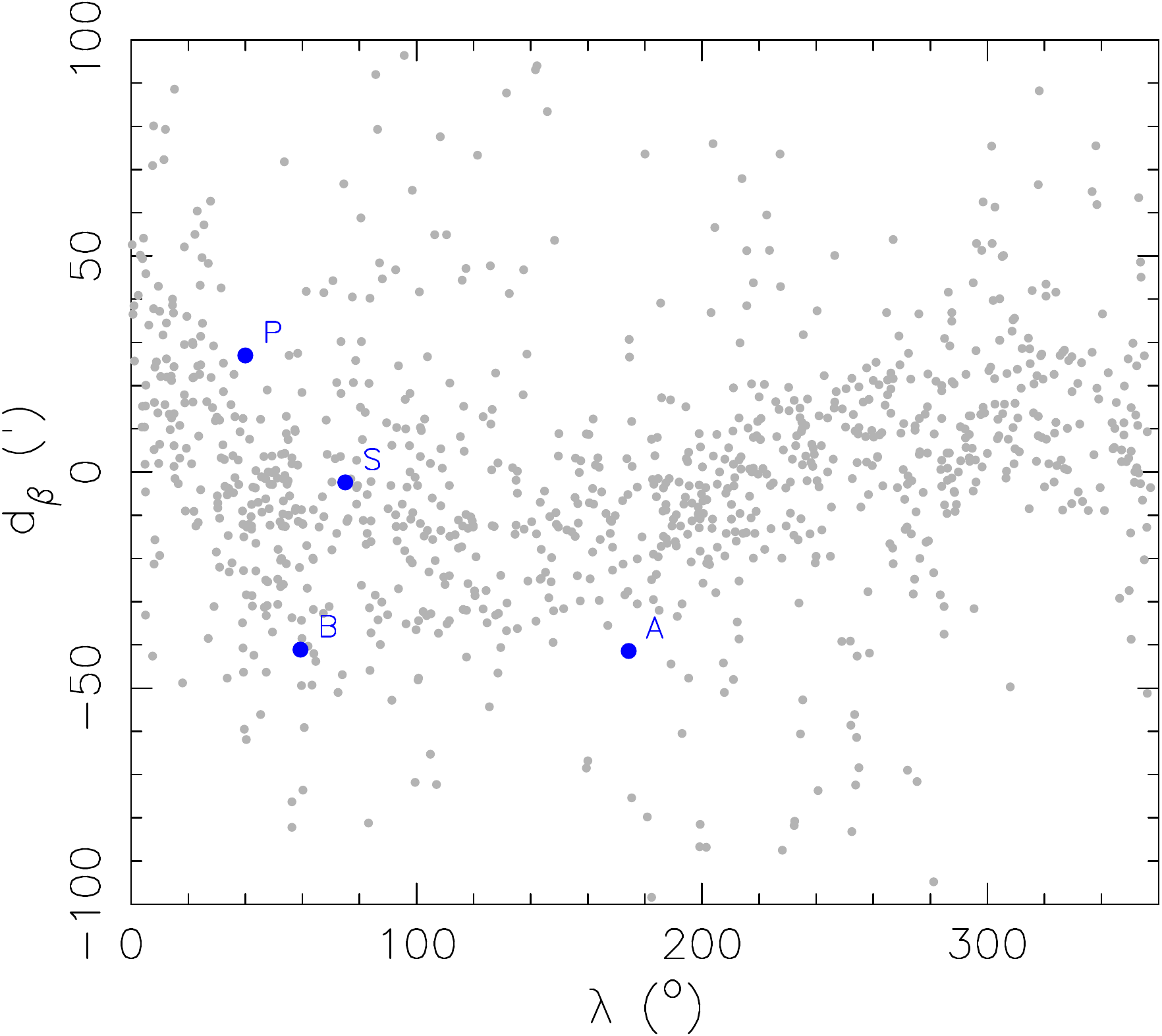}\hfill
\includegraphics[width=0.43\columnwidth]{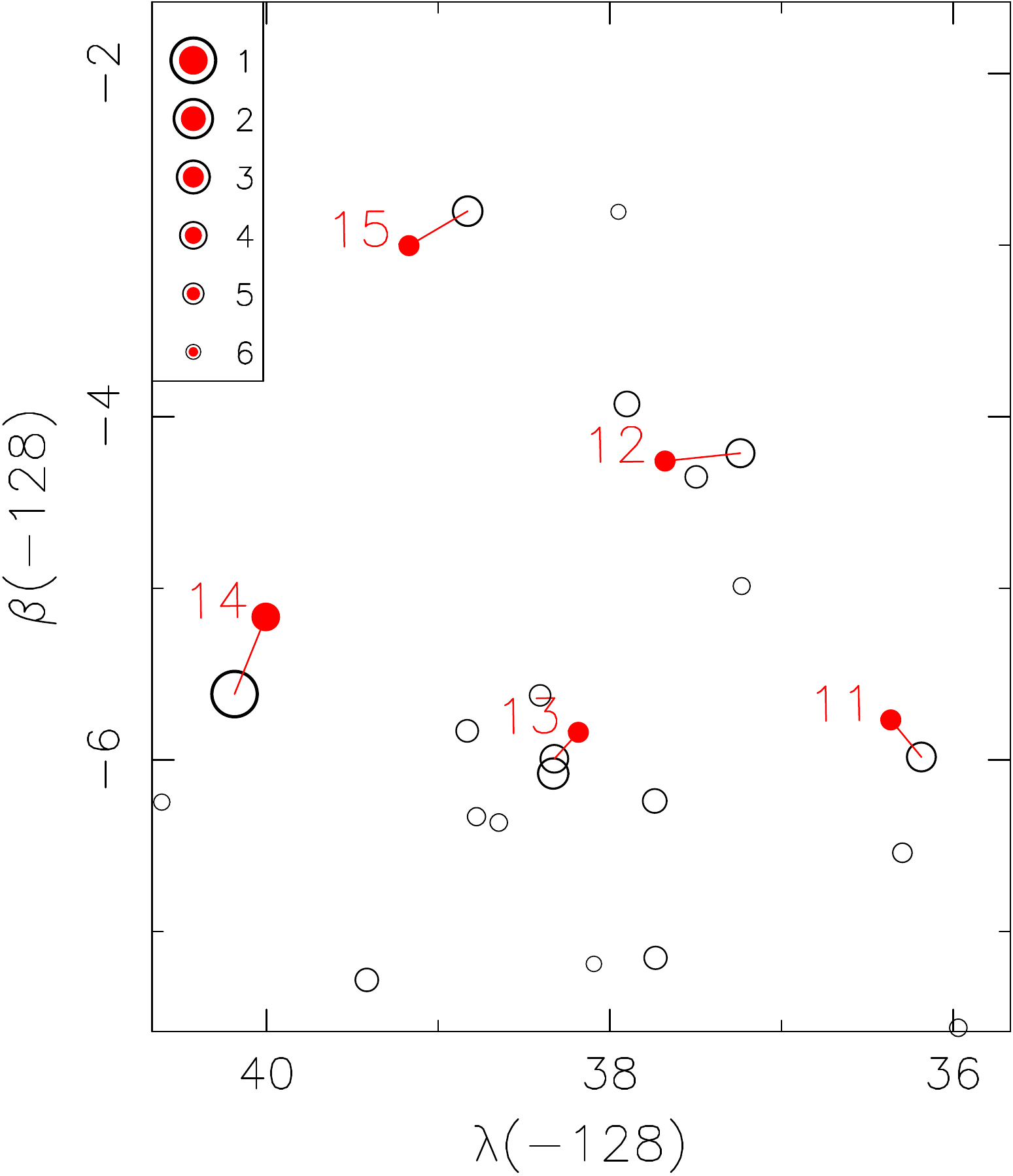}}
\caption{Left: differences $d_\beta$ between correct and catalogued positions of
  stars in Ptolemaios. Stars from Table\,\ref{t:latitudes} are
  highlighted, P=Palilicium (Aldebaran). Right:
  Positions of stars near Aldebaran, no.14 in Taurus in
  the star catalogue of Ptolemaios (red), and correct positions
  computed from data of the HIPPARCOS satellite (black). Scale in
  degrees. The inset indicates the magnitude scale. Ptolemaios puts
  Aldebaran too far North\cite{verbuntgent12}.
\label{f:hyades}}
\end{figure}

\section{Latitude differences: modern}

To see whether the data available to Halley imply proper motion when
analysed with modern methods, we re-analyse them twice, first with the
values for the obliquity used by Halley, and then with the correct
values for the obliquity $\epsilon$.  This enables us to gauge the
effect of using wrong values for $\epsilon$.  We convert the ecliptic
coordinates $\lambda,\beta$ from the Historia
Coelestis\cite{anonymous12} to equatorial coordinates, precess these
with modern equations\cite{seidelman92} to the epoch of Ptolemaios or
Brahe, convert the result into ecliptic coordinates
$\lambda_2,\beta_2$ at the older epoch, and subtract the latitude in
the older catalogue from $\beta_2$.  The equatorial coordinates
computed from the ecliptic coordinates given by Flamsteed are
identical within rounding errors to the equatorial coordinates in his
catalogue, when the value for $\epsilon=23^\circ29'$ is used;
after precession and reconversion to ecliptic coordinates with
$\epsilon=23^\circ51'$, the $\Delta\beta$ listed as
(b) in Table\,\ref{t:pm} may be computed.
A modern equation for the obliquity\cite{seidelman92}
gives $\epsilon=23^\circ28'46.5''$
for 1691 and $\epsilon=23^\circ42'39.6''$ for $-128$;
with these values we compute the $\Delta\beta$ listed as (c) in
Table\,\ref{t:pm}. We also list as (d) the correct change in latitude, 
computed from the data from the HIPPARCOS Catalogue\cite{leeuwen07}.
The same calculation provides the deviation in
latitude $d_\beta$ of the position given by Ptolemaios with 
the correct one. The values for $d_\beta$ in 
Table\,\ref{t:latitudes} are taken from Table 6 of 
Verbunt \&\ van Gent\cite{verbuntgent12} with a change in sign, i.e.\
$d_\beta>0$ indicates that the position given by Ptolemaios is too far
North.  As an illustration we show in Figure\,\ref{f:hyades} the
differences $d_\beta$ for all stars in the Almagest, and also compare
the positions in the Almagest of the stars near Aldebaran with the
correct ones.

Table\,\ref{t:pm} shows that the values we compute with the
obliquities given by Halley are close to the values he gives, but not
identical. Halley indicates that his $\Delta\beta$ for Betelgeuse is a
very rough estimate. For the other stars the difference may arise from
a variety of causes: he certainly used different precession equations,
and possibly used slightly different star positions and obliquity
both for his time and for Ptolemaios, a different epoch for Ptolemaios, and a
different overall correction to the longitudes in Ptolemaios\cite{verbuntgent12}. 
The proximity of the values of $\Delta\beta$ listed as (a) and (b)
in Table\,\ref{t:pm} shows that our effort to replicate Halley is close
to what he actually did.

The effect of a wrong value for the obliquity depends on the
celestial position, in particular on the ecliptic longitude.
It is smallest for Arcturus, and largest for Sirius.

None of the displacements in latitude $\Delta\beta$ computed with
correct values for the obliquities at the epochs of Flamsteed and
Ptolemaios, listed as (c) in Table\,\ref{t:pm}, is significant, due to
the relatively large errors $\sigma_\beta\simeq23'$ in Ptolemaios.
This is true {\em a fortiori} when we realize that Halley did not
study a random selection of stars but {\em selected} some which
appeared to have high proper motion. Our computation using
positions for both epochs derived from HIPPARCOS data gives
the largest change in latitude $\Delta\beta$ between Ptolemaios and
Flamsteed for Arcturus, the star with the smallest $\Delta\beta$ in
the analysis of Halley.  The reason for this is that the error in latitude
by Ptolemaios is in the same direction as the proper motion in
latitude for Arcturus, masking the real change. Conversely, the 
error in latitude by Ptolemaios for Aldebaran is in the opposite direction
as the proper motion in latitude, and adds to the real change.

Turning to the comparison of star positions from Flamsteed and Brahe,
we see that use of the correct value for the obliquity in 1601,
$\epsilon=23^\circ29'28.1''$, leads to a rather smaller displacement
in latitude $\Delta\beta$ for Sirius than the value given by
Halley. With $\sigma_\beta=2'$ for the catalogue of Brahe, this
smaller displacement is not significant.  

We conclude that according to modern criteria, the latitude
differences found by comparing star positions from the epoch of Halley
(presumably from Flamsteed) with those in the catalogues of
Ptolemaios and Brahe provide no evidence for proper motion. 
Halley ignored measurements errors, and this is what led to his 
seemingly positive result.

\section{Occultation of Aldebaran by the Moon: Halley}

\begin{figure}
\centerline{\includegraphics[width=0.8\columnwidth]{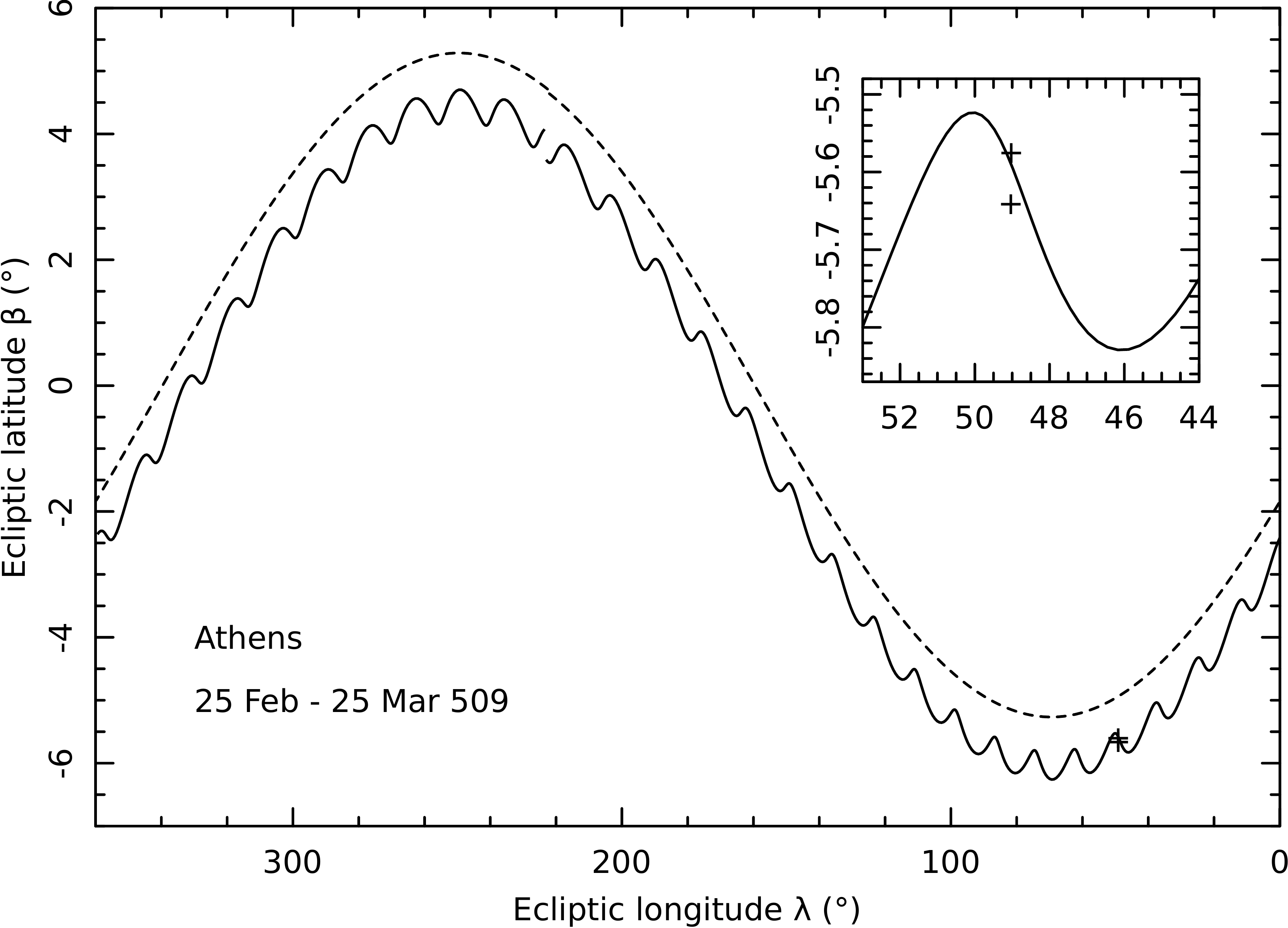}}

\caption{Motion of the Moon along the sky
for the period of a month centered on March 11, 509. The dashed
line gives the geocentric position, i.e.\ the direction of the line
that connects the center of the Earth to the center of
of the Moon. The angle between the lunar orbit and the Earth equator
causes a monthly oscillation. The solid line gives the topocentric position
in Athens of the southernmost point of the Moon, showing the
daily parallax. The upper cross indicates the correct latitude of Aldebaran
in 509 AD, the lower cross the position in 509 computed from the
position in 1690 and zero proper motion. The inset details the motion near 11 March 509.
\label{f:month}}
\end{figure}

Halley's second argument is the occultation of Aldebaran by the Moon
on 11 March 509, described and analysed by
Boulliau\cite{bullialdus45}.  The inclination of the lunar orbit to
the ecliptic causes a monthly oscillation of the position of the Moon
around the ecliptic.  The inclination of the rotation axis of the
Earth with respect to the ecliptic causes a daily oscillation of the
direction to the Moon  as seen from a particular location -- the daily parallax. 
The combined effect of these two oscillations is illustrated for
Athens in Figure\,\ref{f:month}.  The Figure shows that the daily
variation in topocentric latitude of the Moon due to the daily
parallax ($\sim 0.3^\circ$) is much larger than the actual proper
motion of Aldebaran between 509 and 1717 ($\sim 0.065^\circ$).  The
parallax of the Moon depends on its altitude above the horizon, and
hence on (the slowing down of) the rotation of the Earth.  A detailed
knowledge of the Earth's rotation speed is therefore needed to draw
strong conclusions on the proper motion of Aldebaran.  The monthly and
daily variations in the distance to the Moon also cause a monthly and
daily variation in the the apparent size of the Moon.

Whereas Halley was very interested in lunar motion,
his systematic observations of the Moon and analysis of his own and
earlier (in particular Flamsteeds) data only started after 1720. Before
that he computed tables of lunar positions based on Newton's theory.
Whereas these appeared more or less satisfactory for the observations
at hand,  the theoretical predictions worsened rapidly in the next 
18-yr cycle (Cook\cite{cook98}). This implies that
Halley could not compute the position of the Moon on 11 March 509
with sufficient accuracy to decide whether there was an occultation.
The brevity of his report indicates that he did not perform such a
computation -- even if he did, it is clear from modern insight that
the resulting uncertainty was too large to allow significant conclusions.
We return to this below.

It is more likely that Halley based his conclusion on the text by
Boulliau\cite{bullialdus45}. Boulliau prints a Greek text
from the astronomer Heliodorus, and provides a Latin translation. 
We translate from the Greek (see Appendix):
\begin{quote}
On 15 to 16 Phamenoth 225 I saw the Moon following the bright [star]
from the Hyades, after the lighting of the lamps, by at most half a
finger, and it appeared to have occulted it, because the star was next
to the bisection of the convex circumference of the illuminated
part. The true Moon then was at $16^\circ30'$ Taurus.
\end{quote}
The Egyptian date corresponds to 11 March 509, a finger in ancient astronomy
corresponds to 5$'$, and the lamps were lighted after  dusk.
After performing the required computations Boulliau concludes that in
fact no occultation took place. He took 7:20 pm as the local time 
for earliest visibility in Athens,
and used a time difference between Athens and
Hven (Uranienborg) of 45 minutes, close to the correct
value of 44.1 minutes. We list his numbers in Table\,\ref{t:moon}.
Boulliau does not mention a latitude for
Aldebaran, but we may assume that he used the value $-5^\circ31'$
determined in Uranienborg, i.e.\ the value by Brahe (see
Table\,\ref{t:latitudes}). The southern edge of the Moon was taken by
Boulliau to be 15$'$ south of its topocentric center, i.e.\ at
$-5^\circ26'27''$, thus about $4.5'$ north of Aldebaran. Bouilliau
concluded that no occultation could have taken place.

Halley will have noted that the position of Aldebaran
according to Flamsteed is $1'11''$ further North than that
by Brahe (see Table\,\ref{t:latitudes}), but even
then an occultation could only have taken place
if Aldebaran was further north in 509 than this, hence
if Aldebaran moves south with time.

\begin{table}
\small\sf\centering
\caption{Geocentric and topocentric positions of the center of the Moon on 11 March 509 for an observer in 
 Athens according to Boulliau\cite{bullialdus45} (Bo) and according to our modern computations
 for Athens  (At, coordinates  $23^\circ43'40''$ East, $37^\circ59'02''$ North) and 
  for Alexandria (Al, $29^\circ55'$ E, $31^\circ12'$ N).  CA moment of closest approach to Aldebaran;
  EV moment of earliest visibility of Aldebaran. For Athens we also list results for a
  computation in which the length of the day did not change between 509 and 1700,
  hence $\Delta t = \Delta t (1700) = 11$\,s. We also give the correct position of Aldebaran
  in 509, and the position obtained by converting its position in Flamsteed's catalogue
  to 509 without proper motion.
  \label{t:moon}}
\begin{tabular}{|llr|ccccccc|}
\toprule
       & & $\Delta t$ & UT & LT & LST & $\lambda_\mathrm{g}$ &       $\beta_\mathrm{g}$ & $\lambda_\mathrm{t}$ &
       $\beta_\mathrm{t}$\\
       & & (s) & (h) & (h) & (h) & ($^\circ$) & ($^\circ$) &
       ($^\circ$) &   ($^\circ$) \\
\midrule
 Bo & EV &   &  & 19.333 & & 50.507 & $-$4.967 & 49.879 & $-$5.191  \\
\midrule
  At &  CA & 5620 &        13.627 &   15.209 &    2.589 &       48.975 &     $-$4.934 &     48.980 &     $-$5.343\\
  At &  EV & 5620 &        16.733 &   18.315 &    5.703 &       50.552 &     $-$4.982 &     50.025 &     $-$5.271\\
\midrule
  Al &  CA & 5620 &        13.714 &   15.709 &    3.089 &       49.019 &     $-$4.935 &     48.962 &     $-$5.220\\
  Al &  EV & 5620 &        16.333 &   18.328 &    5.715 &       50.349 &     $-$4.976 &     49.808 &     $-$5.160\\
\midrule
   At &  CA &   11 &        16.122 &   17.704 &    5.090 &       49.451 &     $-$4.949 &     49.009 &     $-$5.250\\
\midrule
\multicolumn{8}{|c}{position Aldebaran in 509 computed from HIPPARCOS data} & 49.010 & $-$5.573 \\
\multicolumn{8}{|c}{Flamsteed position Aldebaran converted to 509} & 49.014 & $-$5.641 \\
\bottomrule
\end{tabular}
\end{table}

\section{Occultation of Aldebaran by the Moon: modern}

In our  analysis of the possible occultation of Aldebaran
by the Moon we discuss three problematic aspects of the computation
by Boulliau and -- if he made one independently -- by Halley.
The first is the slow-down of the rotation of the Earth, which was
unknown in the 17th century. This implies that the time elapsed
since 11  March 509 is shorter than one would estimate
based on the length of the day near 1700 AD. Boulliau and Halley
were not aware of this. We investigate this aspect by comparing
correct calculations with those for a constant rotation speed of the Earth.
The second aspect is the question where the occultation
was observed. The Greek text does not mention this, and
Boulliau and Halley assumed that it was in Athens. 
Neugebauer\cite{neugebauer75} argues on the basis of the career
of Heliodorus,  that the observation of the occultation was made in Alexandria.
We investigate this by making calculations both for Athens and for 
Alexandria.
The third aspect is the reliability of the Greek text. We'll discuss
this as we proceed.

We start by computing the relative positions of the Moon and
Aldebaran, for observers in Athens and in Alexandria, on 11 March 509, at two
moments: the moment of closest approach and the moment of
earliest visibility of Aldebaran in the evening. 
To obtain the dynamical time JDE, we add $\Delta t=5620$\,s 
to the Julian Day JD (Morrison \&\
Stephenson\cite{morrison04}, the uncertainty in $\Delta t$ is 140\,s). 

\begin{figure}
\centerline{\includegraphics[width=0.8\columnwidth]{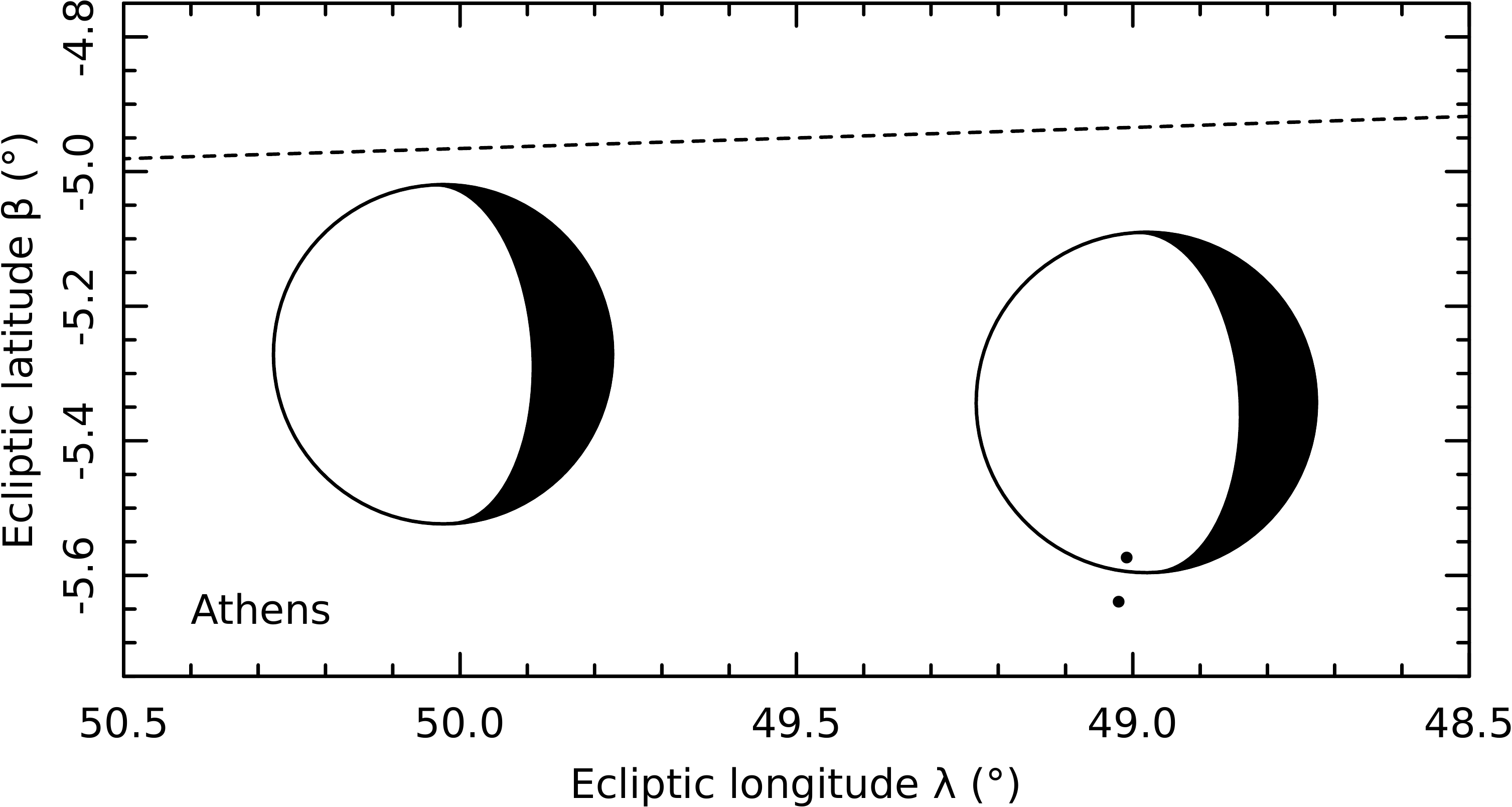}}
\centerline{\includegraphics[width=0.8\columnwidth]{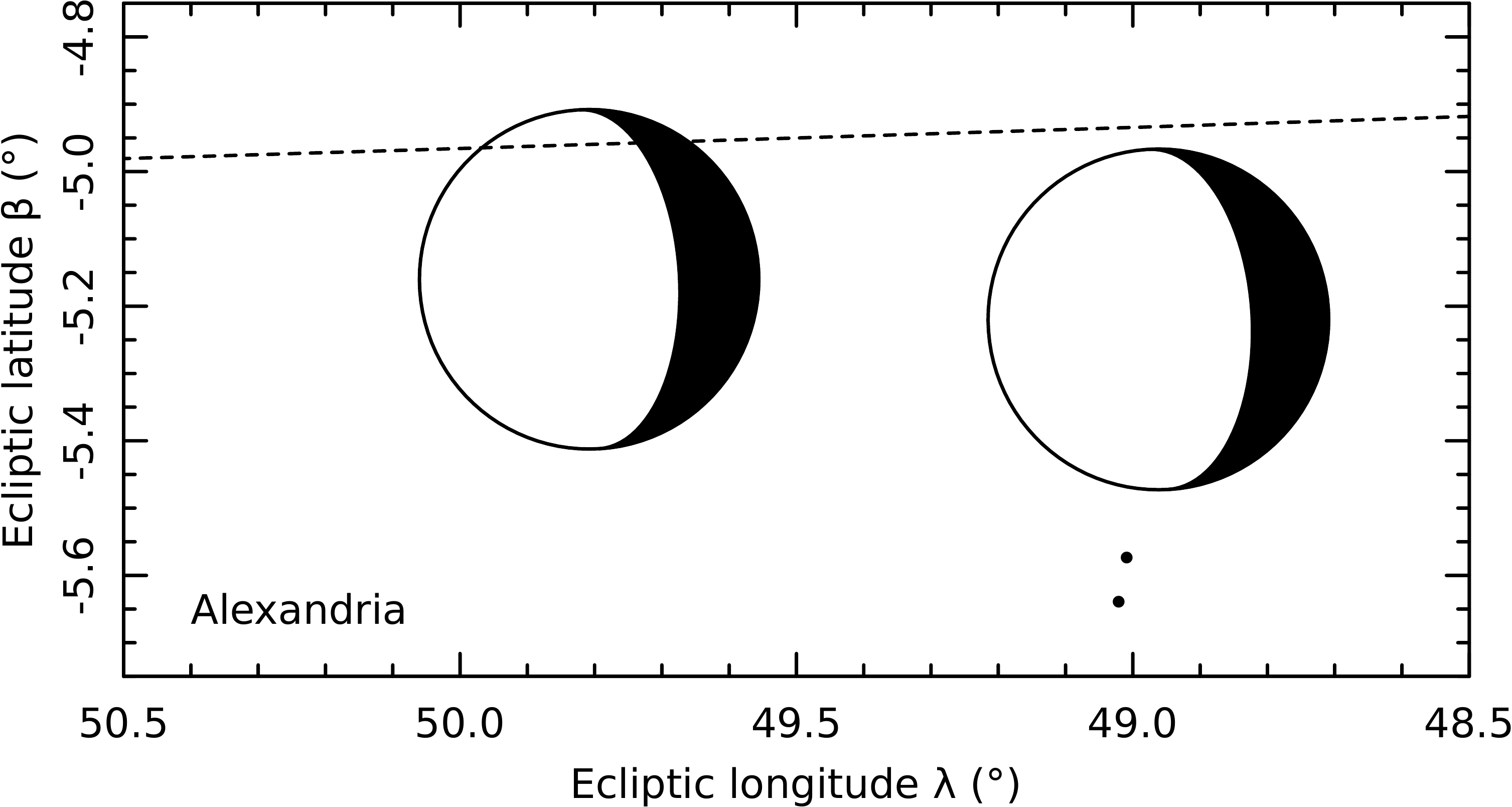}}
\caption{Relative positions of the Moon and Aldebaran at the moment of
  closest approach and at the moment of first visibility of Aldebaran
  for Athens (top) and for Alexandria (below) on 11 March 509,
  computed with modern knowledge. The black crescents correspond
  to the illuminated part of the Moon. The upper position of
  Aldebaran is the correct one, the lower position is computed
  assuming no proper motion between 1690 and 509.  
 The dashed line indicates the geocentric position of the center
 of the Moon \label{f:motion}}
\end{figure}

 We compute the geocentric position and apparent diameter of the
Moon using the fits made to the numerical integration of 
ELP/MPP02\cite{chapront03}.
To find the topocentric position and diameter of the
Moon, we compute the Mean Stellar Time at Greenwich from the dynamic
time, correct it for nutation, and convert it to the Local Stellar
Time at Athens. From this and the geocentric position of the Moon we
compute the daily parallax, and add it to the geocentric position to
obtain the topocentric position and apparent diameter.  Our results
are listed in Table\,\ref{t:moon} and  shown in Figure\,\ref{f:motion}.  
The visible limiting magnitude at the position of Aldebaran depends on its
distance to the horizon, Sun and Moon\cite{schaefer98}.  
We determine the instance when this limiting magnitude equals that of 
Aldebaran  ($V\simeq1$) at dusk on 11 March 509 to be
16:44 UT, which converts to a local time in
Athens of about 6:20 pm. This is the earliest possible moment at
which Aldebaran could theoretically be discerned from the sky background
--- the actual observation was probably somewhat later, and hence the
separation between the Moon and Aldebaran slightly larger. 
We show two positions for Aldebaran:
the correct one computed from HIPPARCOS data\cite{leeuwen07}, and the one computed for
epoch 509 from the position at epoch 1690 in Flamsteed's catalogue
assuming zero proper motion.  In the absence of proper motion, the
ecliptic position of Aldebaran changes due to precession.
The position of the Moon at first visibility indicates a problem
with the Greek text: its edge is not within `a finger', i.e. $<2\farcm5$ from
Aldebaran, but about 46$'$ in Athens and 34$'$ in Alexandria.
If we accept the reading `at most six fingers' from
Heiberg\cite{heiberg07} and Neugebauer\cite{neugebauer75},
the Greek text is more in agreement with an observer in
Alexandria. Even in Alexandria, however, the
position of Aldebaran with respect to the illuminated part of the
Moon does not match the convoluted description in the text.
It is clear that Aldebaran was not occulted on 509 March 11 in Alexandria.
Remarkably, however, we find that the conclusion by Halley is correct
that an occultation of Aldebaran in Athens was possible only
if the star was further North in 509 than in 1690!

It does not follow that Halley proved the proper motion of Aldebaran
with this argument. The accuracy required to prove
or disprove the occultation of Aldebaran in 509 was well beyond reach for
Boulliau, Halley or their contemporaries:  at closest approach Aldebaran was just 1\farcm4
within the limb of the Moon. For example, Halley was not aware
of the slow-down of the rotation of the Earth. If we repeat our
computation for  $\Delta t=11$\,s, ignoring the slowdown of the 
rotation of the Earth between 509 and 1700,  we find a topocentric 
latitude in Athens for the Moon about 6$'$ further North, which implies
there is no occultation of Aldebaran.

\section{Cassini}

Jacques Cassini\cite{cassini38}  in 1738 investigated the possible proper motion of
stars. He notes that Brahe decided that the comparison of
modern measurements of star positions with those made by ancient Greek
astronomers does not provide evidence for proper motion, whereas
Halley decides it does. Cassini concludes that only comparison 
between modern observers can be trusted; as we have seen above, this
conclusion is correct. 

Picard and Jean Dominique Cassini, the father of Jacques Cassini,
made accurate measurements of Arcturus in their study of precession.
Jacques Cassini compares these, and in particular the measurement 
of the ecliptic latitude of Arcturus by Richer in 1672 in Cayenne,
with his own measurement 86 year later in Paris (see Table\,\ref{t:latitudes}), 
and finds a change in latitude of $-2'$.
Cassini remarks that this change is confirmed with
the latitude of Arcturus determined by Flamsteed for 1690.
For comparison with Brahe's measurements, Cassini redetermines the
latitude of Arcturus from an altitude measurement made by Brahe on 
24 February 1584. His better knowledge of refraction,  and especially his 
better value of the obliquity, partially offset
by a less accurate latitude of Hven, enables Cassini to obtain
a more accurate value than Brahe did (Table\,\ref{t:cassini}).
The derived change in latitude between 1584 and 1738 is $5'$ . 
To decide whether this difference is significant, Cassini converts the
meridional altitude $54^\circ36'40''$ of $\eta$\,Boo, measured by
Brahe on 7 February 1586 into a latitude of $28^\circ07'22''$
for $\eta$\,Boo at that date. This is only $3''$ higher than the value
Cassini measures in 1738, and thus, $\eta$\,Boo shows no significant
proper motion. 

\begin{table}
\small\sf\centering
\caption{Ecliptic latitudes used by Cassini\cite{cassini38} to derive the proper motion of Arcturus.
$d$ indicates the difference between the listed value $L$ and the
correct value $C$: $d=L-C$. Italics indicates values derived by us
from other tabulated values. The last column gives values for 1584 from modern computation.
  \label{t:cassini}}
\begin{tabular}{|c|r@{\quad}r@{\quad}rr|r@{\quad}r@{\quad}r|r@{\quad}r@{\quad}r|r@{\quad}r@{\quad}r|}
\toprule
& \multicolumn{13}{|c|}{Arcturus} \\
\midrule
& \multicolumn{4}{|c|}{1672}   & \multicolumn{3}{|c|}{1584 Brahe} &
\multicolumn{3}{|c|}{1584 Brahe-C} & \multicolumn{3}{|c|}{1584 modern} \\
\midrule
& $^\circ$ & $'$ & $"$ & $d(')$  &$^\circ$ & $'$ & $"$  &$^\circ$ & $'$ & $"$ &$^\circ$& $'$ & $"$\\
\midrule
$h$ & & & & &55 & 28 & 15 & 55 & 28 & 15  &  55 & 29 & 25 \\
$R$ & & & & & & & 0 & & & 40 & & & 42 \\
$\phi$ & & & & &  55 & 54 & 30 & 55 & 54 & 15 & 55 & 54 & 28 \\
$\delta$ & 20 & 54 & 15 & $-0.5$ & 21 & 22& 45 & 21 & 21 & 50 & 21 & 23 & 11 \\
$\epsilon$  &23 & 29 & 00& 0.1 & 23 & 31 & 30 & 23 & 29& 30 & 23 & 29 & 36\\
$\alpha$ & {\it 210} & {\it 10} & {\it 45} & 0.0 & & & & 209 & 11 &30 & 209 & 10 &43 \\
$\beta$ & 30 & 57 & 25 & $-0.4$& & &  & 31 & 00 & 29 & 31 & 01 & 26 \\
$\lambda$ & {\it 199} & {\it 39} & {\it 45} &0.2 & & & & {\it 198} & {\it 27} &{\it 28} & 198 & 25 & 58 \\
\bottomrule
\end{tabular}
\end{table}

Cassini proceeds to redetermine from Brahe's altitude measurements the
latitudes of Sirius `near the end of the 16th century' and of Aldebaran
in 1589. From his results, shown in Table\,\ref{t:latitudes}, Cassini
concluded that these stars showed no significant proper motion in
latitude. He drew the same conclusion for other stars from the differences he found
between his own measured latitudes in 1738 and the (redetermined)
latitudes from Brahe's measurements. These differences
(and the values according to modern computations) are 20$''$ for Antares (66$''$), 8$''$ for
$\gamma$\,Aql (53$''$), 22$''$ for Spica (36$''$), 16$''$ for $\alpha$\,CrB (53$''$),
25$''$ for $\alpha$\,Oph (94$''$), 13$''$ for $\alpha$\,Her (59$''$), 
and $<120''$ for Rigel (64$''$), Betelgeuze (66$''$), Regulus (18$''$)
and $\alpha$\,Cap (52$''$).  (Here we assume that the star
`preceding Aquila' is $\gamma$\,Aql.)  Cassini concludes that the
first five have no evidence for proper motion, whereas those with
upper limits of $2'$ are `{\em rather less evident}' than the case for Arcturus.

\begin{table}
\small\sf\centering
\caption{Ecliptic latitudes for Altair and $\beta$\,Aql according to
various catalogues and measurements.\label{t:aquila} }
\begin{tabular}{|l|r@{\quad}r@{\quad}r|r@{\quad}r@{\quad}r@{\quad}r|r@{\quad}r@{\quad}r@{\quad}r|r@{\quad}r@{\quad}r@{\quad}r|}
\toprule
   & \multicolumn{3}{|c|}{Ptolemaios} & \multicolumn{4}{|c|}{Brahe/Cassini} &
       \multicolumn{4}{|c|}{Flamsteed} & \multicolumn{4}{|c|}{Cassini} \\
 & $^\circ$ & $'$ & $d$($'$)  & $^\circ$ & $'$ & $''$ & $d$($'$) 
 & $^\circ$ & $'$ & $''$ & $d$($'$)  & $^\circ$ & $'$ & $''$ & $d$($'$) \\
\midrule
$\alpha$\,Aql & 29 & 10 & $-13.3$ & 29 & 18 & 11 &  $-0.8$
                       & 29 & 19 & 11 & 0.4 & 29 & 19 & 08 & 0.4 \\
$\beta$\,Aql & 27 & 10 & $-0.9$ & 26 & 45 & 08 & $-0.3$ 
                      & 26 & 44 & 20 & 0.4 & 26 & 43 & 40 & 0.4 \\ 
$\Delta\beta$ & 2 & 00 & $-12.4$ & 2 & 33 &03 & $-0.5$ & 
                          2 & 34 & 51 & 0.0 & 2 & 35& 28 & 0.0\\
\bottomrule
\end{tabular}
\end{table}

Finally, Cassini considers the pair $\alpha$\,Aql -- $\beta$\,Aql
(see Table\,\ref{t:aquila}). For $\alpha$\,Aql the latitude has increased
between Ptolemaios and Brahe and on to Flamsteed and Cassini's
measurements, whereas the latitude of $\beta$\,Aql has steadily
decreased over the same time interval.
As a result the difference in latitude between these two stars
has increased by $36'$ since the time of Ptolemaios.

We can see here that Cassini is not consistent in his trust in
numbers from Ptolemaios: where these confirm the trend between
his time and (the revised) Brahe, he accepts them, but when they
do not, as for Aldebaran, he concludes that this `{\em shows that the
  ancient observations are not sufficient for research of this
  type}'. In the cases of Arcturus, Sirius and $\eta$\,Boo he notes
that their latitudes in the catalogue of Ptolemaios are compatible 
with his conclusions from measurements by Brahe and later
astronomers. 

From the above numbers we see that the accuracy of the latitudes
redetermined from Brahe's data by Cassini is generally better than
$1'$, and that the latitudes determined by Richer and Cassini are
generally more accurate than 0.5$'$.
Thus, the proper motion derived by Cassini for Arcturus is
significant, and his conclusion that the data in his possession
do not show significant proper motion for Sirius, Aldebaran,
Betelgeuze and the other stars mentioned above is correct.
In the case of $\beta$\,Aql he is too optimistic: with our 
knowledge of the typical errors in Ptolemaios, Brahe/Cassini,
Flamsteed and Cassini we see from Table\,\ref{t:aquila}
that the differences in its latitude are not significant.

\section{Discussion}

Halley is not the first astronomer who mistook measurement errors
for proper motion.
In comparing contemporaneous observations with earlier
star catalogues, the eight-century Chinese astronomer I-Hsing found 
north-south movements for ten asterisms (Needham\cite{needham59}).
From the magnitude of the displacements, 4 to 5 degrees, it is obvious that,
like Halley, I-Hsing discovered position errors rather than proper
motion.

Occultations of Aldebaran by the Moon in Alexandria do occur, but not on 11 March 509.
An almost central occultation of Aldebaran by the Moon occurred on
12 February 509, about 7 o'clock local time in the morning.
The occultation on 7 April 509 occurred well after Aldebaran became 
visible, after midnight. These occultations do not fit the description
by Heliodorus. If his description fits a real occultation in Alexandria,
it was not one month before or after the date given. In any case,
the text by Heliodorus does not prove proper motion of Aldebaran.

Halley took the latitudes given by ancient astronomers and by
Brahe too much at face value, and as a result interpreted their
measurement errors as evidence for proper motions. In contrast,
Cassini followed Brahe in questioning the reliability of the latitudes given by 
Ptolemaios, and decided they were too uncertain to be
used as evidence for proper motion. By redetermining the
latitudes derived from meridional altitude measurements
by Brahe, Cassini halved the uncertainty in these latitudes,
and thus gave significant proof of the proper motion of
Arcturus, while invalidating the results of Halley.

\section{Acknowledgements}

We thank Dr. Frederik Bakker from the Center for the History of
Philosophy and Science at Radboud University for his invaluable advice
on the interpretation of Latin and Greek texts; and Dr. Robert van Gent for 
pointing us to relevant references.

\section{Notes on contributors}

{\bf Frank Verbunt} works at Radboud University Nijmegen 
in the fields of High-Energy Astrophysics
and Neutron Stars, and  together with Robert van Gent
analysed various historical star catalogues; most recently those
of Ptolemaios and Ulugh Beg$^{13}$

\noindent
{\bf Marc van der Sluys} worked in the fields of Stellar Evolution and
Close Binaries, and Gravitational Waves, also at Radboud University Nijmegen.
He maintains a Dutch website about current astronomical phenomena
{\tt hemel.waarnemen.com}.



\section{Appendix}

The Greek text about the apparent occultation of Aldebaran by the Moon
is printed by Boulliau\cite{bullialdus45} using many ligatures which `more often dismay
than enlighten' the modern reader, to paraphrase Ingram\cite{ingram66}.
We therefore give a transcription kindly provided by Dr.\ Frederik Bakker from the 
Center for the History of Philosophy and Science at Radboud University.

{\selectlanguage{greek}
\noindent ske Famenwj ie e>ic e\stigma, e\iotdac don t`hn sel'hnhn <epom'enhn
\mbox{t\omdac}  lampr\omdac\ t\~wn <U'adwn met`a l'uqnou <af\`hn, <wc
dakt\'ulou, t\`o m\'hkiston <'hmisu, >ed\'okei d\`e ka\`i
>epipro$\sigma$$<$te$>$jhk\'enai \mbox{a>ut\omdacdot} >ep\'eballe g\`ar <o >ast\`hr
\mbox{t\omdac} par\`a t\`hn diqotom\'ian m\'erei t\~hc kurt\~hc
perifere\'ias to\~u pefwtism\'enou m\'erouc, \etadac n d\`e t\'ote <h >akrib\`hs sel\'hnh
per\`i t\`hn i\stigma\ $\mathrm{?}$' mo\~iran to\~u ta\'urou.
}

Here $<${\selectlanguage{greek}te}$>$ is an emendation necessary for the verb form to be
grammatically correct, and the question mark indicates a sign which is
difficult to read in Boulliau. This sign is followed by  $'$ and is
translated by Boulliau as `half' ,which implies that it represents
either $\beta$ or $\angle$, even though in print it looks mostly 
like a {\selectlanguage{greek}\stigma}.
\end{document}